\documentclass[12pt]{article}

\usepackage{graphicx}
\usepackage{epsf}
\usepackage{wrapfig}

\begin{document}

\noindent
{\Large \bf Energy Spectrum of TeV $\gamma$-Rays from Crab Nebula}

\vspace{7 mm}

\noindent
{\large A.K. Konopelko for the HEGRA collaboration}

\vspace{7 mm}

\noindent
{\it Max-Planck-Institut f\"{u}r Kernphysik, Saupfercheckweg 1, 
D-69117 Heidelberg}

\vspace{7 mm}

\noindent
{\bf Abstract} \bigskip

\noindent
The Crab Nebula has been observed by the HEGRA stereoscopic system of 
4 imaging air \v{C}erenkov telescopes (IACTs) for about 100 hours from 1997
September to 1998 March. The recent detailed studies on the system
performance give an energy threshold and an energy
resolution for $\gamma$-rays of 500 GeV and $\sim 18$\%, respectively. 
The Crab spectrum was measured with the IACTs system in a very broad energy 
range from 500 GeV to 25 TeV, using the observations at zenith angles up 
to 50 degree. The pure power-law spectrum in TeV $\gamma$-rays constrains 
the physics parameters of the nebula environment as well as the models of 
photon emission.

\vspace{5 mm}

{\bf 1. Introduction}

The Crab Nebula has been observed and studied over an exclusively 
broad photon energy range embracing  radio, optical, X-ray bands as well 
as high energy $\gamma$-rays up to hundreds of TeVs. The various theoretical 
scenarios of photon emission are primarily based on the Synchrotron
Compton model, which combines the synchrotron and inverse
Compton emissions 
from high-energy electrons, accelerated up to $\sim 100$ TeV, which interact 
with magnetic field and seed photons within the nebula (for review see
Aharonian \& Atoyan, 1998). The predicted spectrum in the TeV energy domain 
appears to be very sensitive to the model parameters: the value of magnetic field, 
the nature of seed photons, the maximum energy of electrons,
etc. Thus a measurement of TeV Crab spectrum sets major constrains on
theoretical expectations.

The imaging air \v{C}erenkov technique was used for
observations in the TeV energy range (for a review see Weekes et al, 1997). 
Currently the Crab serves as a standard candle for TeV $\gamma$-ray
emission. However the Crab energy spectrum, as measured by several groups, still
differs in power-law slope as well as in absolute
normalization. While using the same fundamental principles, 
presently operating imaging air \v{C}erenkov telescopes have a different
energy threshold (e.g. CAT: $\sim 200$ GeV;  Whipple: $\sim 250$ GeV; 
HEGRA: $\sim 500$ GeV; CANGAROO (large zenith angles):$\sim 7$ TeV) and
dynamic energy range. In addition the energy spectrum measurements can be 
affected by the systematic errors, limiting the accuracy of the
result. Stereo imaging gives several advantages for the spectrum 
studies, comparing with the single telescope: (a) direct measurement of 
shower impact parameter; (b) good energy resolution; (c)
wide dynamic range; (d) extended abilities for
systematic studies using several images for an individual shower. 
The HEGRA IACTs system was primarily designed for detailed spectrum 
measurements in the TeV energy domain utilizing the advantages of the
stereoscopic observations.  
Note that detailed systematic studies for the spectrum evaluation technique have been
recently made using Mrk 501 observations in 1997. 
Here we present the Crab data for the last observation period,
analysed with the recently developed technique of energy spectrum
evaluation.

{\bf 2. Observations}

The Crab observations reported here were made with the stereoscopic system 
of 4 HEGRA imaging air \v{C}erenkov telescopes, which are located at La
Palma, Canary Islands (Daum et al., 1997). 
Each of the 4 telescopes consists of a 8.5 $\rm m^2$ 
reflector focussing onto a photomultiplier tube camera. The number of
photomultipliers in the camera was 271, which were arranged in a
hexagonal matrix covering a field of view with a radius of $2.^\circ 3$.  
Any telescope camera was triggered when the signal in two nextneigbour 
of the 271 photomultiplier
tubes exceeded a threshold of 8 photoelectrons, and the system readout 
started when at least two telescopes were triggered by \v{C}erenkov light 
from air shower. The detection rate was 12.6 Hz near the zenith. 
\begin{table}[htb]
Table I. Observation time  \\ \bigskip
\begin{tabular}{lllll}\hline \hline
N & Period & Number of tel-s & Number of runs & Time, hr  \\ \hline
1 & 1997 Sep - 1998 Jan & 4 & 152 & 45.63 \\
2 & 1997 Oct - Nov      & 3 & 48  & 14.26 \\
3 & 1998 Feb - March    & 4 & 84  & 22.62  \\ \hline\hline
\end{tabular}
\end{table}
The Crab was observed in a wobble mode; i.e., the telescopes were
pointed in Declination $\pm 0.5^\circ$ aside from the nominal Crab
position (a sign of angular shift was altered from one run of 20 min to 
another). This is useful for continuous monitoring of the cosmic-ray
background taking the OFF-source region being symmetric about the camera 
center, and $1^\circ$ apart relative to the ON-source
region. Observations of the Crab at zenith angles up to 50 degree were
made from 1997 September 1 to 1998 March 29, for a total of 82.5 hr of 
data taken at good weather. By the fire accident at the HEGRA site, 
one of the telescopes was damaged and was out of operation for a month 
in October-November 1997. At that time, Crab observations were
made with only three telescopes in the system, providing an event
rate of 10.3 Hz near the zenith. Due to unstable weather and
substantial amount of dust which came from Sahara desert to the
island, the average detection rate in 1998 February-March was reduced
down to 10.7 Hz in observations near the zenith. 
The total exposure times for the three periods are summarized in Table 1.

\vspace{4 mm}

{\bf 3. Analysis and result}

The stereoscopic imaging analysis of the data is based on the geometrical
reconstruction of the shower arrival direction and the shower  core
position in the 
observation plane, as well as on the joint parametrization of the shape of the 
\v{C}erenkov light images. The simultaneous registration of several
($\geq$2) \v{C}erenkov light images from air shower provided an
angular resolution of $\sim 0.^\circ 1$ for $\gamma$-ray 
showers. The first $\gamma$-ray selection criterium used here is 
$\rm \Theta^2\leq 0.1\, [deg^2]$, where $\Theta^2$ is the squared angular
distance of the reconstructed shower arrival direction to the source position. 
This orientational cut is very loose and gives a 90\%
$\gamma$-ray acceptance whereas the number of cosmic ray showers is 
reduced by a factor of about 55. In addition we analysed the data by
the ``mean 
scaled width'' parameter,$<\tilde{w}>$. This parameter was introduced in order to
provide an almost constant $\gamma$-ray acceptance over the dynamic energy 
range of the telescope system. Thus, 
the second $\gamma$-ray selection criterium was $<\tilde{w}>\, \leq 1.4$,
which accepts most of the $\gamma$-ray showers ($\sim 99$\%) 
while the relative acceptance of cosmic ray showers is 26\%. The calculation of 
parameter  $<\tilde{w}>$ includes the zenith angle dependence 
of the image shape.  
\begin{table}[htb]
Table II. Crab data summary. \\ \bigskip
\begin{tabular}{llllllll}\hline \hline
Period: & Z.A. & Time, hr & $\rm N_\gamma$ & $\rm N_{cr}$ & $\rm R_\gamma,\,
hr^{-1}$ & $\rm R_{cr}, hr^{-1}$ & S/N$^*$ \\ \hline 
   & $\rm up \, to\,25^\circ$  & 24.9 & 2691 & 5493 & 107 & 220 & 23 \\
1 & $25^\circ - 40^\circ$     & 14.2 & 929   & 3276 & 65   & 230 & 11 \\
   & $40^\circ - 50^\circ$     & 6.5   & 360   & 1355 & 55   & 208 & 7 \\  \hline
   & $\rm up \, to\,25^\circ$  & 9.0   & 788   & 1691 & 88 & 188   & 12 \\
2 & $25^\circ - 40^\circ$     & 3.7   & 235   & 718 & 63 & 194     & 6 \\ 
   & $40^\circ - 50^\circ$    & 1.5    & 28     & 324 & 19 & 216     & 1 \\ \hline
   & $\rm up \, to\,25^\circ$ & 8.5    & 694   & 1584 & 81 & 186 & 11 \\
3 & $25^\circ - 40^\circ$    & 7.7    & 372   & 1533 & 48 & 199 & 6   \\
   & $40^\circ - 50^\circ$    & 6.3    & 197   & 1180 & 31 & 187 & 4
   \\ \hline \hline
\end{tabular} \\ \bigskip
{\small $^*$ - signal-to-noise ratio, $\rm S/N = N_\gamma 
(N_\gamma+2\cdot N_{cr})^{-1/2}, \, N_\gamma = ON-OFF, N_{cr}=OFF$ }
\end{table}
The collection areas, as a function of energy and zenith angle, for
$\gamma$-ray showers has been inferred from Monte Carlo simulations
as described in Konopelko et al., (1998). The energy threshold of
the telescope system, defined as the energy at which the $\gamma$-ray
detection rate reaches its maximum for the differential spectrum $\rm
dN_{\gamma}/dE \sim E^{-2.5}$ , is $\sim 0.5$ TeV at small zenith angles,
and increases up to $\sim2$ TeV at $50^\circ$ zenith angle. The rms
error of the energy determination is $\rm \Delta E/E \sim0.18$.  
\begin{wrapfigure}[19]{l}{8.5 cm}
\epsfxsize= 7.6 cm
\epsffile[6 26 396 405]{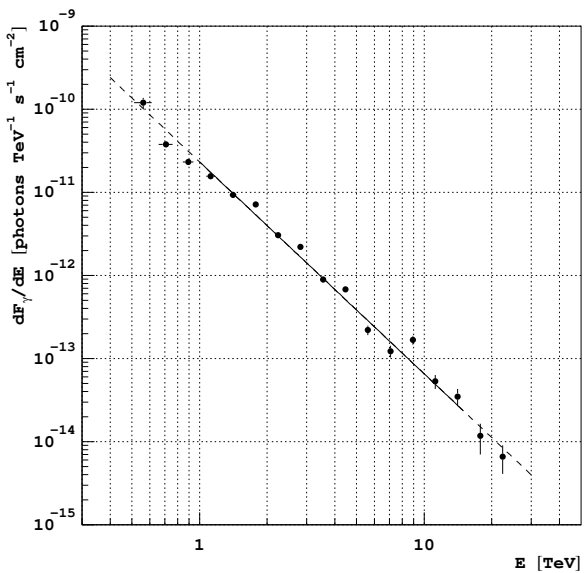} 
\caption{The energy spectrum of Crab Nebula (preliminary).}
 \end{wrapfigure}
In the stereoscopic observations the impact distance of the shower axis 
to a system telescope can be measured with an accuracy $\leq 10$ m. 
The energy of a $\gamma$-ray shower is defined by interpolation over the 
``size'' parameter $\rm S$ (total number of photoelectrons in
image) at the fixed impact distance $R$, as 
$\rm E = f_{MC}( S,R,\theta)$, where $\theta$ is the
zenith angle and $\rm f_{MC}$ is a function obtained from Monte 
Carlo simulations. Note that the Monte Carlo simulations include 
the sampling of detector response in detail. 

The energy distribution for the ON- and OFF-source events, after
the orientation and shape image cuts, were histogramed over the energy 
range from 500 GeV to 30 TeV with 10 bins per decade. The $\gamma$-ray 
energy spectrum was obtained by subtracting ON- and OFF-histograms and
dividing the resulting energy distribution by the corresponding collection 
area and the $\gamma$-ray acceptance. The Crab energy spectrum was then
estimated separately for three observational periods listed in Table
1. Note that Monte Carlo simulations have been adjusted for each of
the periods. As a systematic check of the spectrum evaluation
technique, we compared the energy spectra measured  
from 4 telescopes (period 1) and 3 telescopes (period 2) data 
as well as with the spectra 
evaluated for different zenith angles. These spectra are in a good 
agreement, which  shows the robustness
of the technique.  The preliminary 
Crab energy spectrum, produced by gathering data from three observational 
periods, is shown in Figure 1.
The collection area for $\gamma$-rays rises very fast in the
energy range near the energy threshold of the telescope system, which
is 500 GeV.  Even slight variations of the trigger threshold could
lead to noticeable systematic changes in the predicted spectrum
behaviour in the energy range of $\sim 0.5-1$ TeV. In addition, note
that the statistical significance of detected $\gamma$-rays above 20
TeV  is not very high ($\sim 1.5\sigma$, $\rm N_\gamma$=12). 
Thus in the present analysis we fitted the Crab data in
the energy range 1-20 TeV by a simple power law, which yields 

\begin{equation}
\rm dJ_\gamma (E)/dE =  (2.7\pm 0.2 \pm 0.8) \cdot 10^{-7} {\left(
\frac{E}{1\,TeV}\right)}^{-2.61\pm 0.06\pm 0.1} \,\, photons\, m^{-2}\, s^{-1}
\end{equation}

The first set of errors in eqn.
(1) is statistical and the second systematic is estimated using the Monte Carlo 
simulations which also take into account the current uncertainties in 
absolute calibration of the detector. 
The present Crab flux at 1 TeV, as measured with the HEGRA IACT system, is
higher by a factor of 2 than estimated earlier for the HEGRA CT2 data
using a cosmic ray calibration technique (Konopelko et al, 1996). A more
detailed treatment of the telescope hardware, for example optical smearing, 
increases the CT2 flux to a value (Petry et al, 1996) quite comparable
to the present result.
                      
\vspace{3 mm}

{\bf 4. Discussion}

We derived the spectrum of the Crab Nebula from HEGRA data taken
from 1997 September to 1998 March. The data can be fitted by simple
power law over the energy range 1-20 TeV. It does not exclude a
possible slight steepening of the energy spectral usually predicted by 
inverse Compton modeling of the TeV $\gamma$-ray emission. Even a simple 
power law fit of the Crab data over the energy range 500 GeV to 25 TeV yields a
differential spectrum index of $\sim 2.45$, which indicates a 
flatter spectrum. However, the noted change of the energy spectrum
slope is within the current statistical and systematic errors and the
data overall energy range are consistent with a simple power law fit in 
1-20 TeV given by eqn. (1). Our data matches well at about 20 TeV the recent 
data published by CANGAROO group (Tanimori et al, 1998), which extends up to 
70 TeV. Our data are consistent with a flat power law index of $\sim 2.5$ 
beyond 10 TeV, which is likely to be for TeV $\gamma$-rays from $\pi^0$s, 
generated by protons accelerated at the termination shock of the pulsar wind. 

\vspace{3 mm}

{\bf References}

\noindent
Aharonian, F., Atoyan, A. 1998, {\it Neutron Star \& Pulsars,} UAP, Tokyo, 439  

\noindent
Daum et al. 1997, {\it Astroparticle Phys.} 8, N1-2, 1 

\noindent
Konopelko et al. 1996, {\it Astroparticle Phys.}, 4, 199

\noindent
Konopelko et al. 1998, submitted to {\it Astroparticle Phys.}

\noindent
Petry et al. 1996, {\it A \& A}, 311, L13

\noindent
Tanimori, T. et al. 1998, {\it ApJ}, 492, L33 

\noindent
Weekes et al. 1997,{\it Proc. 4th Compton Sym.}, Part I, Williamsburg, 361

\end{document}